# The use of ChatGPT in higher education: The advantages and disadvantages

Joshua Ebere Chukwuere

North-West University, South Africa

joshchukwuere@gmail.com

**Abstract**

Higher education scholars are interested in an artificial intelligence (AI) technology called ChatGPT, which was developed by OpenAI. Whether ChatGPT can improve learning is still a topic of debate among experts. This concise overview of the literature examines the application of ChatGPT in higher education to comprehend and produce high-level instruction. By examining the essential literature, this study seeks to provide a thorough assessment of the advantages and disadvantages of utilizing ChatGPT in higher education settings. But it's crucial to consider both the positive and negative elements. For this rapid review, the researcher searched Google Scholar, Scopus, and others between January 2023 and July 2023 for prior research from various publications. These studies were examined. The study found that employing ChatGPT in higher education is beneficial for a number of reasons. It can provide individualized instruction, and prompt feedback, facilitate access to learning, and promote student interaction. These benefits could improve the learning environment and make it more fun for academics and students. The cons of ChatGPT are equally present. These problems include the inability to comprehend emotions, the lack of social interaction chances, technological limitations, and the dangers of depending too much on ChatGPT for higher education. Higher education should combine ChatGPT with other teaching techniques to provide students and lecturers with a comprehensive education. However, it is crucial to consider the positives, negatives, and moral issues before adopting ChatGPT in the classroom.

**Keywords:** ChatGPT, Chatbots, OpenAI, natural language processing (NLP), Higher education, Technology, Artificial Intelligence (AI)

## Introduction

The artificial intelligence (AI) revolution, which is bringing about significant change, includes chatbots. In many facets of life, including education, chatbots are rising in popularity. Higher education institutions (HEIs) are starting to utilize chatbots that can comprehend student concerns, offer assistance, and deliver answers (Nalyvaiko & Maliutina, 2021). Chatbots will enhance support services, academic counseling, and student participation through the power of natural language processing (NLP). However, how much NLP can aid higher education is yet unknown. Higher education has recently increased its use of technology. These include sophisticated language models like ChatGPT. ChatGPT method have provided scholars with fresh approaches to teaching students one-on-one and increasing their involvement in their education. The GPT-3 was used by OpenAI to construct the ChatGPT large language model.

For speaking and studying in higher education, the ChatGPT AI language model created by OpenAI (Taecharungroj, 2023; Rasul, Nair, Kalendra, Robin, de Oliveira Santini, Ladeira & Heathcote, 2023; Meyer, Urbanowicz, Martin, O'Connor, Li, Peng & Moore, 2023) is ideal. It can make text that sounds like a person talking and can reply well to many different things. People are increasingly interested in and talking about using ChatGPT in HEIs. Some academics are worried about the issues with ChatGPT. Some people believe that it doesn't understand feelings very well, doesn't have real human contact, and may cause scholars and students to depend too heavily on technology.

It is important to consider the advantages and disadvantages of using ChatGPT in HEIs, and how it can be effectively used in education. This article explains how ChatGPT might impact higher education in the future, mentioning both positive and negative effects. This research also tells experts how it can assist scholars and students in acquiring knowledge on new topics, and what might occur as a consequence, and provides suggestions and ideas for future studies on this subject. The researcher in this study wants to explain how ChatGPT can be used in HEIs to make learning better for people who read it.

**Background**

Automated assistance called chatbots are getting more and more popular because they can give customers quick and helpful support. HEIs have also started using this technology to improve the learning and teaching experiences for their students and scholars. Chatbots can help with things like finding information about courses, organizing schedules, and giving overall advice. They can use natural language processing (NLP) to provide customized learning experiences that are based on what students like and what they need (Fuchs, 2023; Caratozzolo, Rodriguez-Ruiz & Alvarez-Delgado, 2022). Chatbots through ChatGPT and other similar applications (tools) are always available to help with lots of questions. This makes them a good and easy option for students who need help even when lecturers (educators) are not available or the office is closed.

ChatGPT is a smart computer program that can chat with people across disciplines including students and educators. It uses advanced technology to understand and respond to questions and conversations. Created by OpenAI and using the GPT-3.5 system, it can give short or detailed answers to a variety of questions and requests (Zhang, Zhang, Li, Qiao, Zheng, Dam & Hong et al., 2023; Willems, 2023). For HEIs, ChatGPT is a useful tool that can understand and respond to difficult questions, which makes it a great option for higher education. However, it's necessary to think about the possible disadvantages of ChatGPT along with their advantages before using them in HEIs support services. Although ChatGPT is helpful, it should be used with other teaching methods to give students a complete education.

**History of ChatGPT**

ChatGPT is a computer program that can interact with a person. It runs on OpenAI, which is a group that does research. The company started in 2015 with Sam Altman, Greg Brockman, Elon Musk, and some other people. The Transformer model, created by OpenAI in 2017, is a kind of computer program that uses a neural network. It is the basic starting point for ChatGPT. Compared to older models, the Transformer was specifically designed to better understand and analyze sequential information, such as language (Vaswani, Shazeer, Parmar, Uszkoreit, Jones, Gomez & Polosukhin, 2017).

Launched on November 30, 2022 (Zhai, 2022), ChatGPT quickly gained over one million subscribers as people took to social media to spread the word about its potential (Kay, 2023). Generative Pre-trained Transformer, or GPT, was initially introduced by OpenAI in 2018. GPT was developed to generate original text in response to a given prompt. It is programmed and trained on a large corpus of text. GPT, however, is unable to generate graphics or images. The development of subsequent versions of the model, culminating in the launch of GPT-3 in June 2020, was driven by the model's ability to generate coherent, substantial, and realistic text. With over 175 billion parameters, GPT-3, among the latest version of ChatGPT, is currently the most robust and powerful. Numerous applications, such as chatbots, natural language processing, and language translation, have utilized it. The ability of GPT-3 to perform a wide range of language tasks, such as translation, summarization, and question-answering, without the need for task-specific training data, is one of the most significant advancements of this technology (Brown, Mann, Ryder, Subbiah, Kaplan, Dhariwal & Amodei, 2020).

However, there have been several issues with the development of ChatGPT. Concerns regarding the potential misuse of language models like GPT-3 have been raised by several researchers. They are worried about the spread of biased or harmful information, as well as the generation of fake news. Others have expressed ethical concerns regarding the lack of transparency in the development of these models and the potential for them to perpetuate existing biases in linguistic data (Yan, Fauss, Hao & Cui, 2023). Despite these reservations, ChatGPT is still a crucial tool for natural language processing. It has the power to completely change how people interact with linguistic data and provide speedy answers to inquiries (questions). This rapid review article demonstrates the enormous potential power of AI through ChatGPT, and any other emerging technologies such as machine learning, and deep learning in higher education. It also demonstrates how these technologies are continuously improving and evolving in facilitating the teaching and learning process.

**Research method**

To quickly evaluate the volume of literature on the use and effectiveness of ChatGPT in the context of higher education, the researcher combined focused keyword searches with a rigorous manual screening of significant academic articles. The initial assessment entails the identification of key phrases such as "History of ChatGPT," "Background," "advantages and disadvantages," and "ChatGPT in Higher Education." Utilizing a rapid and effective way to locate and gather pertinent research on utilizing ChatGPT in higher education, the study was promptly reviewed. Finding 15 significant publications required the researcher to look through recent materials and seek certain terms. Journal, conference, and book articles that employed various techniques to obtain information, such as counting and measuring as well as screening, were included in these publications. The purpose of the rapid review was to summarise the study's principal findings. It considered ChatGPT's advantages and disadvantages and made recommendations on how it may be improved going forward. The researcher collected information from past studies to give a complete evaluation of what is currently known about the subject.

The scope of the rapid review was constrained to works published within the past five years particularly work published between January 2023 and July 2023, potentially resulting in the exclusion of pertinent material published before this timeframe. The utilization of the rapid review approach enabled the researcher to efficiently select and integrate the primary findings from various studies, thereby providing a valuable overview of the present research landscape concerning the utilization and applications of ChatGPT in higher education institutions. In general, the rapid review method proved to be effective in efficiently summarizing the body of research on the utilization of ChatGPT in higher education. While systematic reviews offer a comprehensive and structured evaluation of the literature, rapid reviews are unable to achieve the same level of thoroughness and rigor. Future research in this domain should prioritize the resolution of these challenges and the exploration of optimal strategies for the effective integration of ChatGPT in higher education.

**The critics of the use of ChatGPT in higher education**

Despite the potential to revolutionize the student experience in higher education, there are several concerns associated with the utilization of chatbots such as ChatGPT. The absence of emotional intelligence in chatbots is a significant criticism, as it may lead to frustration and dissatisfaction among users (Prioleau, 2020). Students may opt to engage in a conversation with a live advisor (educator) who can provide personalized and empathetic support. Another criticism directed towards ChatGPT pertains to their potential difficulty in comprehending complex inquiries, leading to potential misunderstandings. Despite the fact that chatbots such as ChatGPT are designed to search, organize, comprehend, and respond to a wide range of questions and inquiries, they may not be capable of providing the same level of nuanced expertise as a human advisor (Zielinski, Winker, Aggarwal, Ferris, Heinemann, Lapeña & Citrome, 2023).

Inquiring about information regarding unfamiliar organizations or notable individuals can be considered a personal question or inquiry, for instance. Taecharungroj (2023) asserts that a significant number of online users express criticism towards the accuracy, bias, and inadequate performance of ChatGPT replies. Furthermore, ChatGPT has garnered criticism from numerous sources due to its potential to generate harmful content and disseminate false information. Furthermore, it possesses the capability to diminish critical thinking and human cognition, while also causing users to become inattentive and unproductive.

Some people are worried that chatbots like ChatGPT might take over the jobs of human advisers, which could result in job losses and a decrease in the quality of support services (Amundsen & Johansen, 2022; Prioleau, 2020). Chatbots can help reduce the workload for teachers and staff, but universities should be careful not to rely too heavily on them as the main way of helping students. Overall, using chatbots like ChatGPT in university can make the student experience better. However, before opting to implement ChatGPT in their support of academic services, higher education institutions should carefully evaluate any potential issues and ChatGPT's potential limits.

**Advantages of ChatGPT in higher education**

According to empirical studies, using ChatGPT increases students' happiness and engagement in studying. According to research by Kasneci, Seßler, Küchemann, Bannert, Dementieva, Fischer and Kasneci (2023), ChatGPT can encourage student engagement in their coursework and research by providing prompt responses and tailored guidance. According to research by

Javaid, Haleem, Singh, Khan and Khan (2023), ChatGPT can lessen the workload that faculty and staff members have to bear. This gives them more time to concentrate on other crucial responsibilities. However, there are a number of advantages to using ChatGPT in higher education:

- **Enhanced student engagement:** Students can benefit from ChatGPT by having quick access to knowledge and assistance anytime they require it. Students may use this feature to communicate with the chatbot anytime they want, which enables them to obtain the support and guidance they want. Additionally, ChatGPT may provide responses that are customized to the individual requirements and interests of every student. This keeps students interested in and involved in what they are studying. Students learning is made more enjoyable with chatbots like ChatGPT. They support students in continuing to take an active role in their education. According to research by Dwivedi, Kshetri, Hughes, Slade, Jeyaraj, Kar and Wright (2023), these sophisticated ChatGPTs provide students with an engaging and dynamic approach to learning.
- **Cost-effective:** The benefit of ChatGPT is that it may address several inquiries (questions) at once, reducing the workload on staff. Organizations save time and money by doing this. With ChatGPT, you may quickly get assistance with admissions, course registration, obtaining financial aid, and academic counseling. This strategy enables staff members to devote more time to challenging jobs.
- **Instant feedback:** Students can immediately correct their errors. Ray (2023) claims that ChatGPT can provide responses and feedback on student work rapidly. This aids in their learning and allows them to correct their errors.
- **Availability:** Students can use ChatGPT to get immediate answers to their questions whenever they need them. This helps them to quickly get the information they need. If students are better at making decisions, it will make their learning in school better.
- **Scalability:** ChatGPT can handle a lot of inquiries as the institution grows, making it very efficient. ChatGPT can help support services by learning to handle more types of questions as more students come.
- **Personalized learning:** ChatGPT can provide customized learning that meets the specific needs and interests of each student. They can also give suggestions on what lessons to take, give feedback, and provide study materials. Being able to understand what each student needs and giving them individualized help allows for a more customized way of learning. (Fushs, 2023).
- **Multilingual support:** ChatGPT can help schools with many different students because it can understand and speak different languages well. ChatGPT can understand and answer in many languages, making it easier for international students to get information and help.
- **Improved accessibility:** Make it easier for students with disabilities to access by using ChatGPT. They can help students who can't see well or hear well access and understand written information. ChatGPT can help make education more accessible for students who don't speak English as their first language or have disabilities. Some claim that utilizing specialized technology to assist students with impairments or translate languages in real-time might be beneficial (Kasneci et al., 2023).

- **Data collection:** ChatGPT can compile data on student interactions and interests. This data may be utilized to personalize learning experiences and raise the level of support given. Additionally, this data may be utilized to identify certain areas where students might require special assistance.

**Disadvantages of ChatGPT in higher education**

Numerous concerns and problems have been raised by the usage and integration of chatbots in higher education institutions, such as ChatGPT. Chatbots, according to their critics, lack the emotional intelligence that is essential for human interaction and communication, which might cause user frustration and dissatisfaction. Furthermore, ChatGPT may struggle to understand complex queries, potentially leading to misunderstandings. The disadvantages of ChatGPT in higher education may include:

- **Lack of human touch:** Their interactions demonstrate emotional intelligence. If students only use a chatbot as their primary mode of contact, they may feel detached from the institution, which may reduce their overall satisfaction. According to Bozkurt, Xiao, Lambert, Pazurek, Crompton, Koseoglu, and Jandri (2023), the lack of in-person interaction and the limited personal engagement given by ChatGPT may lead to feelings of alienation and decreased motivation in students.
- **Limited flexibility:** Because they are designed to react to certain questions, chatbots can be unable to offer individualized responses that call for contextual understanding. Students requiring more intricate assistance may experience feelings of frustration and dissatisfaction as a consequence.
- **Limited accuracy:** Students may experience frustration and confusion due to the chatbot's provision of erroneous or inadequate information. The consequences for students, including course selection, financial assistance, and academic guidance, might suffer from inaccurate information.
- **Security concerns:** Chatbots are vulnerable to hacking and online attacks, potentially compromising the confidentiality of student data. The security of ChatGPT and the protection of student data are the responsibility of the institutions.
- **Limited emotional intelligence:** When addressing student difficulties (personal, social, economic, and political challenges), chatbots might lack sympathy, empathy, and understanding since they lack the emotional intelligence that comes from human involvement and connection. Students who believe their problems are not being effectively handled may get frustrated and dissatisfied as a result of this. Students may experience frustration or a sense of alienation since the chatbot (ChatGPT) lacks emotional intelligence and the capacity to recognize and respond to students' emotional states properly (Sallam, Salim, Barakat & Al-Tammemi, 2023; Tatsenko & Donik, 2023).
- **Technical limitations:** Chatbots rely on imperfect natural language processing technologies. They could have trouble comprehending difficult questions or knowing how to answer particular kinds of inquiries. Students may get perplexed and frustrated as a result of this.

- **Maintenance and upkeep:** For chatbots to be functional, frequent maintenance and upkeep are necessary. For institutions, this can be time-consuming and expensive, especially if they lack specialized IT employees to operate the chatbot.
- **Risk of bias:** Chatbots may be designed with prejudice or stereotypes that might be harmful to particular student groups. The programming of ChatGPT should prioritize inclusivity and take into account the diverse student populations, as stated by educational institutions.
- **Over-reliance on technology:** Overuse of ChatGPT and other technologies can have detrimental effects on critical thinking abilities and problem-solving inventiveness (Qadir, 2023; Božić, n.d).

If higher education institutions want to use ChatGPT in their support teaching and learning, they must carefully weigh these benefits and drawbacks.

**The Contributions, Implications, and Recommendations of the study**

Students' ability to approach a problem, learn, and engage with educational information has the potential to be disrupted and revolutionized by the usage and use of ChatGPT in higher education. It does, however, also provide some restrictions and difficulties that must be taken into account. The following are some of the contributions, implications, and suggestions made concerning the subject.

**Contributions**

By offering individualized and prompt feedback, boosting accessibility and responsiveness, and raising student engagement, the usage and use of ChatGPT in higher education has the ability to disrupt and revolutionize how students perceive educational content problems and learn. It can provide students with a more engaging and customized learning experience, improving academic achievement and overall learning satisfaction. ChatGPT has the ability to improve the whole teaching experience by easing the burden on educators and allowing them to devote more time to various elements of education, such as increasing student participation and collaboration. The usage and application of ChatGPT can help lecturers devote more time to teaching and learning, administrative duties, community engagement, and other intellectual tasks.

**Implications**

Higher education is significantly impacted by ChatGPT because it has the ability to completely change how students engage with and learn from instructional resources. It is crucial to understand that there are some restrictions and dangers associated with this occurrence. There may be a dearth of emotional intelligence, a reduction in interpersonal interaction, and an over-dependence on technology. It has the power to foster an atmosphere where employees, including supervisors and lecturers, are more prone to engage in irrational thought and analysis. Higher education administration, educators (lecturers), and students must carefully consider the consequences of adopting ChatGPT at a higher education institution in order to ensure its successful and responsible usage. A further finding is that in order to successfully integrate

ChatGPT into the educational process, lecturers (educators) need to have the right training and assistance.

**Recommendations**

- **Integration with other teaching strategies:** ChatGPT should be used in conjunction with other teaching methods including group discussions and peer-to-peer interactions to provide a more complete learning experience.
- **Training data quality:** The quality, standard, and accuracy of the training data should be carefully chosen, evaluated, and updated in order to ensure that ChatGPT is accurate, objective, and yields the right results.
- **Professional development for educators:** To properly use ChatGPT in their instructional practices and support students in overcoming the limitations and challenges associated with the technology. Therefore, educators must undergo professional development and training to acquire the necessary skills and knowledge.
- **Ongoing evaluation:** A continuing study of the utilization and effectiveness of ChatGPT in higher education institutions in order to ascertain its efficacy and identify areas for further development.

ChatGPT has the potential to enhance the teaching and learning process in higher education by introducing fresh chances for individualized instruction and immediate feedback. However, it also has several drawbacks and difficulties that need to be properly evaluated and resolved. Academic institutions have the potential to effectively integrate this technology into the educational process and promote student success through the adoption of a well-rounded approach that considers the benefits, drawbacks, and recommendations associated with ChatGPT in higher education.

**Future studies**

Future studies on the use and applications of ChatGPT in higher education should include a wide variety of issues. Investigating ChatGPT's efficacy in certain subject areas, such as science, technology, engineering, and mathematics (STEM), where individualized education and immediate feedback are critical, is one option. This might include comparing the academic performance of students who use ChatGPT against those of students who do not.

A good next step may be to examine how ChatGPT supports collaborative learning. To raise the standard of online conversations and team projects, ChatGPT provides real-time comments and suggestions. Future research may examine ChatGPT's effects on group participation and the standard of collaborative work. Future studies could look at how ChatGPT might help and direct students with their mental health. Students can utilize ChatGPT as a private, secure setting where they can speak honestly about their issues and get emotional support. For children who lack access to conventional counseling services, this resource might be of great help.

The shortcomings of ChatGPT, particularly its lack of emotional intelligence and interpersonal interactions, should be the focus of future studies. To enhance the educational experience, researchers may look at combining ChatGPT with other teaching strategies including group discussions and peer-to-peer exchanges. Future studies on ChatGPT in higher education might provide insightful information on how to use this technology to enhance learning outcomes and address problems that the field of education is currently dealing with.

**Conclusion**

By giving lecturers (educators) students continuous access to information and support services, chatbots like ChatGPT have the potential to revolutionize the higher education sector. Although there are certain limitations to adopting ChatGPT, the advantages such as higher student engagement, cost-effectiveness, accessibility, and scalability far outweigh the risks. The safety, reliability, and compatibility of chatbots with student needs must be guaranteed when educational institutions look at using them in the classroom. Institutions may improve students' academic experiences by offering a more personalized and interesting learning environment by incorporating artificial intelligence (AI) and chatbots. This approach aims to provide pupils with the knowledge and abilities they'll need to be successful in their future activities.